# SIMULATION OF ARSENIC DIFFUSION DURING RAPID THERMAL ANNEALING OF SILICON LAYERS DOPED WITH LOW-ENERGY HIGH-DOSE ION IMPLANTATION

O.I. Velichko[1], A.M. Mironov[2], V.A. Tsurko[3], G.M. Zayats[3]

[1]Belarusian State University on Informatics and Radioelectronics, 6 P. Brovka Street, Minsk, 220013 Belarus; e-mail: oleg_velichko@lycos.com

[2]Institute of Applied Physics Problems, Belarusian State University, 7 Kurchatov Street, Minsk, 220064 Belarus; e-mail: MironovA@bsu.by

[3]institute of Mathematics, Academy of Sciences of Belarus, 11 Surganova Street, Minsk, 220072 Belarus; e-mail: vtsurko@im.bas-net.by

The model of transient enhanced diffusion of ion-implanted As is formulated and the finite-difference method for numerical solution of the system of equations obtained is developed. The nonuniform distribution of point defects near the interface and more accurate description of arsenic clustering are simultaneously taking into account. Simulation of As diffusion during rapid annealing gives a reasonable agreement with the experimental data.

## Introduction

As the lateral dimensions of modern integrated circuits are scaled down to the submicrometer range, the need for accurate models of silicon doping is increased. The well-known and widely used models of transient enhanced diffusion during rapid thermal annealing of semiconductor substrates very often lead to the results that disagree with the experimental data for low-energy high-dose ion implantation. The difference is mainly due to using inadequate clustering models for describing high concentration dopant diffusion [1] and due to the influence of interfaces on the distributions of defects [2].

In this paper the model of transient enhanced diffusion of ion-implanted arsenic simultaneously taking into account the nonuniform distribution of point defects near the interface and more accurate consideration of arsenic clustering is presented. To solve the system of diffusion equations obtained, a numerical algorithm was formulated and a code was developed. Simulation of the diffusion of ion-implanted arsenic in the vicinity of the surface agrees well with the experimental data of [3].

## Model

It is suggested that arsenic diffusion occurs due formation, migration, and dissociation of the pairs "$As^+ D^r$", where $As^+$ and $D^r$ are the substitutionally dissolved arsenic atom and intrinsic point defect, respectively [4]. A thermodynamic approach based on the local equilibrium between the substitutionally dissolved arsenic, point defects and the pairs leads to the following system of diffusion equations:

**Equation of diffusion of dopant atoms**

$$\frac{\partial C^T}{\partial t} = \sum_{k=1}^{p} \frac{\partial}{\partial x_k} D \left[ \frac{\partial (\tilde{C} C)}{\partial x} + \frac{\tilde{C} C}{\chi} \frac{\partial \chi}{\partial x_k} \right]. \quad (1)$$

**Equation of diffusion-drift-reaction of point defects** [5]

$$\sum_{k=1}^{p} \frac{\partial}{\partial x_k} \left[ d^C \frac{\partial \tilde{C}}{\partial x_k} - \frac{\mathrm{v}}{d_i} \tilde{C} \right] - \frac{k^C \tilde{C}}{l_i^{\times 2}} + \frac{\tilde{C}^g}{l_i^{\times 2}} = 0. \quad (2)$$

Here

$$C^T = C + C^{AC},$$

$$D = D_i D^C(\chi),$$

$$\tilde{C} = C^\times / C_i^\times, \qquad d^C(\chi) = d(\chi)/d_i,$$

$$D^C(\chi) = \frac{1 + \beta_1 \chi + \beta_2 \chi^2}{1 + \beta_1 + \beta_2},$$

$$D_i = D_i^\times + D_i^1 + D_i^2,$$

$$\beta_k = \frac{D_i^k}{D_i^\times},$$

where $C$ and $C^{AC}$ are the concentrations of substitutionally dissolved arsenic atoms and dopant atoms incorporated into clusters, respectively; $D = D(\chi)$ and $D_i$ are the effective and intrinsic diffusivities of arsenic, respectively; $D_i^\times$, $D_i^1$, and $D_i^2$ are the partial diffusion coefficients due to interactions of the dopant atoms with the neutral, singly, and doubly charged defects, respectively; $\chi$ is the concentration of electrons normalized to the intrinsic carrier concentration $n_i$; $C^\times$ is the concentration of point defects in the neutral charge state; $C_i^\times$ is the equilibrium concentration of neutral point defects in the bulk of the semiconductor; $d = d(\chi)$ and $d_i$ are the effective and intrinsic diffusivities of point defects governing the arsenic diffusion; $k^C = k^C(\chi)$ and $l_i^\times$ are the effective absorption coefficient and average migration length of point defects, respectively; $v$ is the effective drift velocity of mobile defects due to elastic stress; $\tilde{C}^g$ is the effective generation rate of point defects normalized to the rate of equilibrium thermal generation [5].

To calculate concentration of clustered arsenic atoms $C^{AC}$, a new model of clustering [1] is used. It was shown in [1] that assumption of the formation of doubly negatively charged clusters $(VAs_2)^{2-}$ which incorporated two arsenic atoms led to a better fit to the experimental data. Then the concentration of clustered arsenic atoms can be obtained from the following expression:

$$C^{AC} = K\tilde{C}_D \chi^4 C^2,$$

where $K$ is the characteristic parameter of clustering; $\tilde{C}_D$ is the normalized concentration of the defects participating in cluster formation. To calculate $\chi$, a condition of local charge neutrality can be used:

$$\chi = n/n_i = \frac{C - C^{AC} - C^B + \sqrt{(C - C^{AC} - C^B)^2 + 4n_i^2}}{2n_i},$$

where $C^B$ is the summarized concentration of acceptors.

**Numerical method**

Let us define the mesh $\omega_\tau = \{t_j = j \cdot \tau, j = 0,1,...,j_0, j_0\tau = t_0\}$ in the domain $[0 \le t \le t_A]$ and the mesh $\omega_h = \{x_1^{(i_1)} = x_{i_1} = i_1 h_1, i_1 = 0,1,...,N_1, h_1 N_1 = l_1,$

$x_2^{(i_2)} = x_{i_2} = i_2 h_2, i_2 = 0,1,...,N_2, h_2 N_2 = l_2\}$ in the space domain G.

To find the approximate solution of the system of equations (1) and (2) on the mesh $\omega_\tau \times \omega_h$, the finite-difference method [6] is used. Following [6], we approximate Eqs. (1) and (2) by the system of nonlinear algebraic equations

$$\left(y + Kz^4 y^2\right)_{\bar{t}}\Big|_{i_1,i_2} =$$

$$= \sum_{k=1}^{p}\left(a_k(z)\bar{y}_{\bar{x}_k} + \frac{\bar{y}}{z}z_{\bar{x}_k}\right)_{x_k}^{j}\Big|_{i_1,i_2}, \quad p = 1,2, \quad (3)$$

$j = 1,2,...,j_0, i_1 = 1,2,...,N_1 - 1, i_2 = 1,2,...,N_2 - 1,$

$$z = \frac{1}{2n_i}(y - Kz^4 y^2 - N^B +$$

$$+ \sqrt{(y - Kz^4 y^2 - N^B)^2 + 4n_i^2}\Big|_{i_1,i_2}^{j}, \quad (4)$$

$j = 0,1,...,j_0, i_1 = 0,1,...,N_1, i_2 = 0,1,...,N_2,$

$$\sum_{k=1}^{p}\tilde{y}_{\bar{x}x_k} + b_1 \tilde{y}_{\overset{\circ}{x_k}} + b_2 \tilde{y} + b_3\Big|_{i_1,i_2} = 0, \quad (5)$$

$i_1 = 1,2,...,N_1 - 1, i_2 = 1,2,...,N_2 - 1.$

Here $z$ is the approximate value of $\chi$; $\bar{y} = \tilde{y}y$; $\tilde{y}$ and $y$ are the approximate values of $\tilde{C}$ and $C$, respectively. Let us

$$a_1(z(x_{i_1}, x_{i_1}, t_j)) =$$

$$= \frac{1}{2}\left(D(z(x_{i_1}, x_{i_1}, t_j)) + D(z(x_{i_1-1}, x_{i_1}, t_j))\right)$$

$i_1 = 1,2,...,N_1, i_2 = 1,2,...,N_2, j = 1,2,...,j_0,$

$$a_2(z(x_{i_1}, x_{i_1}, t_j)) =$$

$$= \frac{1}{2}\left(D(z(x_{i_1}, x_{i_1}, t_j)) + D(z(x_{i_1}, x_{i_1-1}, t_j))\right)$$

$i_1 = 1,2,...,N_1, i_2 = 1,2,...,N_2, j = 1,2,...,j_0.$

In Eqs. (3), (4), and (5) we used the well-known designations [6] for a difference approximation of the derivatives. To obtain a numerical solution, corresponding boundary conditions are to be added to the algebraic system (3), (4), and (5). For example, we used reflection boundary conditions in the calculations presented below. The solution of Eq. (3) is obtained by the direct scalar sweep method for the case of $p=1$ and direct matrix sweep method for $p=2$ (see, for example [7]).

The initial values for the mesh functions $y$ and $z$ were obtained from the values of calculated or measured arsenic distribution after ion implantation.

Using the obtained values of $\tilde{y}$, we find the values of $y$ and $z$ from the system of equations (3) and (4) by the method of iterations.

### Results of simulation

The calculated profiles of total As concentration and concentration of substitutionally dissolved arsenic atoms after annealing are presented in Fig. 1. The calculated time-average distribution of point defects in the neutral charge state during annealing is also presented. For comparison with the experimental data, the diffusion process investigated in [3] was simulated. In [3], (100)-oriented, p-type silicon substrates of 15 Ω·cm nominal resistivity were heavily implanted with As at 35 keV to a dose of 5×10$^{15}$ ions/cm$^2$ in a random direction. The annealing was carried out at a temperature of 1030 $^0$C for 5 sec in a nitrogen atmosphere. The arsenic concentration profile after annealing was measured by secondary ion mass spetrometry and presented in Fig. 1 by circles.

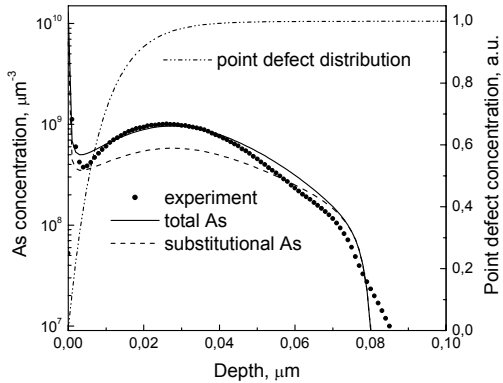

Fig. 1: Calculated As profiles after annealing of the ion-implanted layer. The time-average profile of point defects during annealing is presented by the dash-dotted line. Experimental data are taken from [3].

It can be seen from Fig. 1 that the calculated profile of the total As concentration agrees well with the experimental data, including the near surface region. Consequently, the proposed model allows simulation of high concentration transient enhanced diffusion of As implanted in Si, although this transport process differs substantially from the diffusion processes described by Fick's second law.

We use the following values of the parameters that describe diffusion of the dopant atoms and point defects: $D_i \tilde{C}_m$ = 1.0×10$^{-6}$ μm$^2$/sec; $n_i$ = 1.04×10$^7$ μm$^{-3}$; $\beta_1$ = 3.67; $\beta_2$ = 0.34; $k^{AC}$ = 3.4×10$^{-15}$ μm$^3$; $\tilde{C}_s/\tilde{C}_m$ = 0.002; $l_i^{\times}$ =0.008 μm. Here $\tilde{C}_s$ and $\tilde{C}_m$ are the time averaged values of the relative defect concentration on the surface and beyond the doped layer, respectively. The values of the equilibrium arsenic diffusivity at 1030 $^0$C known from the literature lie in the range $D_i$ = 3.98×10$^{-7}$ — 8.59×10$^{-7}$ μm$^2$/sec. It means that the dopant diffusion is enhanced approximately from 2.5 to 1.16 times due to the nonequilibrium point defects $\tilde{C}_m$. It is important to note that nonuniform distribution of point defects plays the main role in the "uphill" diffusion of arsenic atoms near the surface of the semiconductor and formation of distinct local maximum of the dopant concentration during annealing.

### Conclusions

A model of high concentration transient enhanced diffusion of As implanted into Si has been developed. The model takes into account the nonuniform distribution of point defects near the interface and provides more accurate consideration of arsenic clustering. These features make it possible to use the model for simulating As diffusion during rapid thermal annealing of the layers doped low-energy high-dose ion implantation. The results of numerical calculations of As redistribution during thermal annealing agree well with the experimental data, which confirms the adequacy of the model proposed.